\documentclass{ws-procs9x6} 
\begin{document} 
\title{ MASSES AND BOOST-INVARIANT WAVE FUNCTIONS OF \\ 
        HEAVY QUARKONIA FROM \\ 
        THE LIGHT-FRONT HAMILTONIAN OF QCD} 
\author{ Stanis{\l}aw D. G{\l}azek } 
\address{ Institute of Theoretical Physics, Warsaw University,\\ 
          Ho\.za 69, 00-681 Warsaw, Poland\\ 
          $^*$E-mail: stglazek@fuw.edu.pl\\
          www.fuw.edu.pl/$\sim$stglazek } 
\begin{abstract} 
A new scheme for calculating masses and boost-invariant wave
functions of heavy quarkonia is developed in a light-front 
Hamiltonian formulation of QCD. Only the simplest approximate 
version with one flavor of quarks and an ansatz for the mass 
gap for gluons is discussed. The resulting spectra look 
reasonably good in view of the crude approximations made in 
the simplest version.
\end{abstract}
\bodymatter 
\section{Motivation for the LF Hamiltonian approach to QCD} 
\label{Motivation} 
The method for calculating masses and wave functions of
heavy quarkonia that is reported here stems from the program
of a weak-coupling expansion for Hamiltonians in light-front
(LF) QCD\cite{LFQCD}. The LF form of dynamics was discovered
by Dirac\cite{Dirac1,Dirac2} and continues to excite
imagination of physicists\cite{Wilson2004}. Many authors
have rediscovered LF dynamics. A famous example concerns
application to hard exclusive processes\cite{frontreview0}.
Review articles provide other references concerning LF
formalism\cite{frontreview1,frontreview2}. One reason of the
great interest is that the field quantization on the front
hyperplane leads to 7 kinematical generators of the
Poincar\'e group, instead of only 6 kinematical generators
in the standard form (3 momentum and 3 angular momentum
operators). Another reason is that the vacuum problem in the
LF formulation of quantum field theory appears intriguingly
different from the standard version. The same two reasons
propelled also the development of the renormalization group
procedure for effective particles (RGPEP) that is the basis
of the method discussed here\cite{RGPEP}. But there are two
more reasons.

The first is that the LF Fock space of free bare particles
can be introduced before one constructs the concept of a
quantum field operator and builds Hamiltonian interaction
terms for the bare quanta using such field
operators\cite{Weinberg}. This is useful when one attempts
to mathematically define a theory of quarks and gluons that
never appear as incoming or outgoing particles in scattering
experiments but exist inside hadrons. Proceeding in this
order, one can regulate the interaction terms in the LF
Hamiltonian in a boost-invariant way. The regularization is
accomplished using the relative transverse momenta and
fractions of total "plus" momentum that the bare particles
in interaction are carrying\cite{RGPEP}. The transverse and
"plus" momenta are defined with respect to the direction of
the front hyperplane, the latter conventionally defined by
the condition $x^+ = x^0 + x^3=0$ in a frame of reference in
which the front is moving along $z$-axis ($x^3$), extending
in the transverse directions of coordinates $x^\perp$. The
transverse momenta of the particles are denoted by $k^\perp$
and their "plus" momenta by $k^+ = k^0 + k^3$. The regulated
interaction Hamiltonian for bare particles is invariant with
respect to boosts along the $z$-axis and two additional
boost-like transformations that can change transverse
momenta to arbitrary values. It is also invariant with
respect to two translations in the $\perp$ directions,
translation in the $x^-$ direction, and rotations around the
$z$-axis (typically directed along the beam, a dominant
momentum transfer, or a suitable combination thereof
depending on a scattering experiment, but in a complete
theory the choice should not matter). Thus, the Hamiltonian
has the same structure in a large class of frames of
reference (7 dimensional). Consequently, one does not need
to construct Hamiltonian counterterms that restore boost
symmetry when one tries to quantitatively explain the
mechanism by which masses, spins, and other quantum numbers
of hadrons are formed. Most attractively, the basic
Hamiltonian has the same structure in the rest frame of a
hadron, where the constituent picture works\cite{PDG}, and
in the infinite momentum frame, where the parton model
works\cite{partons}. The LF Hamiltonian approach raises hopes
for conceptual and quantitative explanation of the
constituent and parton models in a single and complete
formulation of QCD.

The second reason is that one can take advantage of the
concept of potentials acting at a distance between
relativistic quarks and gluons, a feast not conceivable in
the standard approach that is defined using objects
distributed on a space-like hyperplane in space-time. Every
interaction between two objects located at different points
of such a hyperplane corresponds to a dynamical effect
spreading faster than light and has to eventually cancel out
in observables (this happens in perturbative QED but it is
not clear how it may happen in non-perturbative QCD).
Therefore, it is common in the standard approach to consider
only action and use local Lagrangian densities for field
variables in a path-integral formula for transition
amplitudes. Geometrical ideas such as strings and other
nonlocal objects in multidimensional spaces are then used to
regulate and explain the interaction terms in the
Lagrangians. An additional argument for the Lagrangian
approach is that it can incorporate variation of the metric
in space-time and, hopefully, illuminate the problem of
connection between particle dynamics and
gravity\cite{Strings}. But if one leaves gravity aside as
too weak to be of an immediate concern at the scale of
hadronic binding mechanism, it is useful to observe that the
LF Hamiltonian at $x^+=0$ can contain potential terms that
act between particles separated by arbitrarily large
distances and such interactions can obey the rule that
dynamical effects do not spread faster than light. Namely,
when the bare point-like particles have the same transverse
positions, the four-dimensional space-time interval between
them is zero no matter how large is their separation in the
direction of $x^-$. In fact, the LF counterpart of the
Coulomb interaction between two particles 1 and 2 on the LF
is proportional to $|x_1^- - x_2^-|$ when $x_1^\perp =
x_2^\perp$, and otherwise vanishes. Precisely this type of
interaction leads to a model of confinement in a 1+1
dimensional theory\cite{tHooft}. It is clear that the LF
Hamiltonians are very singular when transverse distances
between charged point-like particles tend to zero, and the
singular terms can involve entire functions of the $x^-$
distances between the particles. 

Both reasons described above indicate that one needs a powerful 
ultraviolet renormalization technique for Hamiltonians in order 
to develop LF QCD (note that the Wilsonian concept of universality 
could help in identifying effective Hamiltonians irrespectively of 
many details in setting up the initial bare theory). A new technique 
has been invented\cite{similarity1,similarity2} and adopted in a 
general scheme of weak coupling expansion in LF QCD\cite{LFQCD}. 
More recently, the Hamiltonian approach has been redesigned in 
the form of RGPEP\cite{RGPEP}. The results reported below are 
obtained using RGPEP and an ansatz for a mass-gap for gluons. 
\section{ Binding above threshold in heavy quarkonia }
\label{Binding}
Since there is not enough room here to thoroughly explain
the RGPEP in application to heavy quarkonia\cite{ho} in
comparison to other approaches, only the main steps are
indicated. The central puzzle is how a systematic treatment
of QCD can produce binding of quarks above threshold. QED
describes quantum binding only below the mass threshold. 
So, how can binding above the threshold emerge in a 
relativistic quantum theory? 

One begins from the Lagrangian for QCD with one flavor of
quarks
\begin{eqnarray} 
{\cal L} = \bar \psi(i\hspace{-4pt}\not\!\!D - m)\psi - \frac{1}{4}F^{\mu\nu a}F_{\mu\nu}^a \, . 
\end{eqnarray} 
A canonical LF procedure in gauge $A^+ = 0$ produces a
Hamiltonian with many terms (constraint equations are solved explicitly)\cite{frontreview0}
\begin{eqnarray} 
H_{can} &=&
H_{\psi^2} + H_{A^2} + H_{\psi A \psi} + H_{(\psi\psi)^2}
\nonumber \\
& + & 
H_{A^3} + H_{A^4} +  H_{\psi A A \psi} + H_{[\partial A A]^2} + H_{[\partial A A](\psi\psi)}  \, .
\end{eqnarray} 
Each of these terms is an integral of the corresponding Hamiltonian density
over the LF hyperplane with $x^+=0$, $H = \int dx^- d^2x^\perp {\cal H}$. 
For example, 
\begin{eqnarray} 
{\cal H}_{\psi^2} 
&=&
\frac{1}{2} \bar \psi \gamma^+
\frac{-\partial^{\perp \, 2} + m^2 }{
i\partial^+} \psi \, , 
\quad \quad \quad 
{\cal H}_{A^2} =
- \frac{1}{2} A^\perp (\partial^\perp)^2 A^\perp \, ,  \\ 
{\cal H}_{\psi A \psi} 
&=& g
\, \bar \psi \hspace{-4pt}\not\!\!A \psi \, , 
\quad \quad \quad 
{\cal H}_{(\psi\psi)^2} = \frac{1}{2}g^2
\, \bar \psi \gamma^+ t^a \psi \frac{1}{(i\partial^+)^2 } \bar \psi \gamma^+ t^a \psi \, , \,\, etc. 
\end{eqnarray} 
The fields at $x^+ = 0$ are expanded into creation and
annihilation operators for bare quarks and gluons, the 
measure is $[k] = dk^+ d^2k^\perp /(16 \pi^3 k^+)$:
\begin{eqnarray}
\label{Psi} 
\psi = \sum_{\sigma c} \int [k] \left[
\chi_c u_{k\sigma} b_{k\sigma c} e^{-ikx} + \chi_c
v_{k\sigma} d^\dagger_{k\sigma c} e^{ikx} \right] \, ,  \\
\label{A} 
A^\mu = \sum_{\sigma c} \int [k] \left[ t^c
\varepsilon^\mu_{k\sigma} a_{k\sigma c} e^{-ikx} + t^c
\varepsilon^{\mu *}_{k\sigma} a^\dagger_{k\sigma c}
e^{ikx}\right] \, ,
\end{eqnarray}
$c$ stands for color, $\sigma$ for spin. The bilinear terms 
in $\cal H$ provide kinetic energies for the bare
particles and the terms with more than two fields provide
interactions that are regulated\cite{ho} as indicated in
Section \ref{Motivation}. The regularization implies
appearance of counterterms, $H_{CT}$, that restore the
dynamics that was cut off by the regularization. The full
regulated Hamiltonian, $H = \left[ H_{can} + H_{CT}
\right]_{reg}$ provides the initial condition for RGPEP
(RGPEP is also used to determine $H_{CT}$)\cite{RGPEP}. The
main step is to replace the canonical operators $b$, $d$,
and $a$, or their hermitean conjugates in Eqs. (\ref{Psi})
and (\ref{A}), commonly denoted by $q_{can}$, by unitarily
equivalent operators that create or annihilate effective
particles corresponding to the renormalization group
parameter $\lambda$, $q_\lambda = U_\lambda \, q_{can} \,
U_\lambda^\dagger$, so that $q_\infty = q_{can}$ and
$dH_\lambda/d\lambda = [{\cal T}_\lambda, H_\lambda ]$,
where ${\cal T}_\lambda = dU_\lambda/d\lambda \,
U^\dagger_\lambda$. Given the initial condition $H_\infty =
\left[ H_{can} + H_{CT} \right]_{reg}$, one can
systematically evaluate the Hamiltonian $H_\lambda =
H_\infty + \int_\infty^\lambda ds [{\cal T}_s , H_s ]$ in
perturbation theory. $H_\lambda$ is equal to $H$ but it is
expressed in terms of operators creating and annihilating
effective particles of size $1/\lambda$ with respect to
strong interactions. Since $H_\lambda$ is expressed in terms
of the creation and annihilation operators for effective
quarks and gluons instead of the bare canonical ones, it
contains different interaction terms, including new
effective potentials. For $\lambda$ on the order of hadronic
masses, the effective particles are expected to correspond to 
the constituent quarks and gluons that are used to describe
hadrons in particle tables\cite{PDG}. The perturbative
procedure for evaluating $H_\lambda$ is safe from genuine 
infrared singularities because the RGPEP generator 
${\cal T}_\lambda$ is designed to exclude the possibility 
that energy denominators in perturbation theory are 
significantly smaller than $\lambda$. 

A quarkonium eigenvalue problem for the QCD Hamiltonian $H_\lambda$, 
$(P^+ H_\lambda - P^{\perp \,2})|P\rangle = M^2 |P\rangle$, 
is solved by first eliminating the eigenvalues $P^+$ and 
$P^\perp$ of three kinematical momentum operators $P^+_\lambda$ 
and $P^\perp_\lambda$ (these operators are also provided by 
RGPEP\cite{algebra}) and obtaining an eigenvalue equation for the
quarkonium mass $M$ (the center-of-mass motion is eliminated 
from the eigenvalue problem exactly). Still, the eigenstate 
$|P\rangle$ is built from the virtual effective particles
in the LF Fock space and carries four-momentum $P$ with $P^-= 
(M^2 + P^{\perp \, 2})/P^+$. In terms of the effective
quark-antiquark, quark-antiquark-gluon, and other components:
\begin{eqnarray}
|P\rangle   & = & |Q_\lambda \bar Q_\lambda \rangle + 
            |Q_\lambda \bar Q_\lambda g_\lambda \rangle + 
            \, . \, . \, . \, \, . 
\end{eqnarray}
This expansion may converge, in distinction from the
expansion of the same state into canonical bare-particle
sectors, because interactions in $H_\lambda$ are 
limited to momentum transfers smaller than $\lambda$ by the
form factors $f_\lambda$ that appear in all interaction vertices
in $H_\lambda$. The form factors are introduced through the generator 
${\cal T}_\lambda$ of RGPEP. In the effective-particle basis, 
the Hamiltonian $H_\lambda$ takes a matrix form 
\begin{eqnarray}
\left[H_\lambda\right] & = & 
\left[
\begin{array}{ccc}
\cdot~~~      &  \cdot    ~  &  \cdot     \\
\cdot~~~      &  T_3 + V_3~  &  Y         \\
\cdot~~~      &  Y^\dagger~  &  T_2 + V_2 
\end{array}
\right]
\rightarrow 
H_{2+3}=
\left[
\begin{array}{cc}
T_3 + \mu^2_{ansatz} &  Y  \\
Y^\dagger   &  T_2 + V_2
\end{array}
\right] \, ,  
\end{eqnarray}
in which dots denote couplings with sectors with more than 3
effective constituents, $T$ refers to kinetic energy terms,
$V$ to potentials, $Y$ to emission of effective gluons by
quarks, and 2 and 3 to the Fock components with 2 and 3
effective particles. The arrow indicates a truncation of the
system to sectors $|Q_\lambda \bar Q_\lambda \rangle$ and
$|Q_\lambda \bar Q_\lambda g_\lambda \rangle$ only, which is 
done at the price of introducing an ansatz for the gluon mass
gap,
\begin{eqnarray}
\mu^2_{ansatz} & = & \left( 1 - \frac{\alpha^2}{\alpha_s^2}
\right) \, \mu^2 \, .
\end{eqnarray}
The ansatz is so designed that when the coupling constant $\alpha$ 
(this is the effective coupling at some small scale $\lambda$) 
is extrapolated to a realistically strong value $\alpha_s$, the 
ansatz will be removed and the true QCD interactions can be 
recovered order-by-order in the weak coupling expansion in 
$\alpha$. The gap function $\mu^2$ is inserted in order to model 
the effect of all the non-abelian Coulomb potentials, $V_3$, that act in 
sector $|Q_\lambda \bar Q_\lambda g_\lambda \rangle$  and the 
interactions that produce couplings to additional sectors with 
more constituents (the dots). It is very unlikely that the first 
approximation in QCD should be $\mu^2=0$. But if $\mu^2 \neq 0$, 
the resulting dynamics in the $|Q_\lambda \bar Q_\lambda \rangle$ 
sector is described in the leading order in $\alpha$ by the eigenvalue 
equation  $H_{Q \bar Q \lambda} |P\rangle = M^2 |P\rangle$,
where the effective quark-antiquark Hamiltonian has the form (qualitatively) 
\begin{eqnarray}
\label{ansatzscheme2}
H_{Q\bar Q \lambda } & = & 
T_2 + V_2 + 
Y^\dagger  \frac{1}{T_3 + \mu^2}  Y \, . 
\end{eqnarray}
The main point is that the gluon emission and absorption 
produces diverging (for small $q_z$) terms of the form
$f_\lambda \, \frac{ 4m^2 }{ q_z^2 } \, \frac{\mu^2 }{ q^2 +
\mu^2 } \, f_\lambda$, in which the momentum transfer 
$\vec q$ approaches zero. This happens also in the quark 
self-interaction terms. The net effect is positive, lifting 
the quark energy above threshold. In addition, the factor 
dependent on $\mu^2$ becomes 1 for small $\vec q$ irrespectively 
of the details of the ansatz for $\mu^2$. The final result\cite{ho} 
is a harmonic oscillator potential that appears as a leading 
correction to the color Coulomb interaction at typical distances
between the quarks (the Coulomb term appears with the Breit-Fermi 
spin factors). Technically, it is the harmonic oscillator term 
that leads to the binding above threshold, $M > 2m$, where $m$ 
is the mass ascribed to the quarks. Such
effect is absent in positronium in QED because there are no 
Coulomb-like interactions between photons and electrons and 
no mass-gap for photons .
\section{ Masses and wave functions in the crudest approximation }
The resulting eigenvalue equation for quarkonium wave
function can be solved numerically and the mass spectrum
depends on the choice of the coupling constant $\alpha$ and 
quark mass $m$ at some value of $\lambda$. The Breit-Fermi
terms include three-dimensional $\delta$-functions that are
smeared and made finite by the presence of the form factors
$f_\lambda$. If one assumes $\alpha_{M_Z} \sim 0.12$, the 
RGPEP evolution with one flavor of quarks in the same Hamiltonian
scheme\cite{glambda} gives $\alpha \sim 0.326$ at 
$\lambda \sim 3.7$ GeV. Table \ref{Table} shows masses of
$b \bar b$ quarkonia obtained for $\alpha$ = 0.326 and 
$m$ = 4857 MeV at $\lambda$ = 3699 MeV, adjusted
to fit masses of $\chi_1$(1P) and $\chi_1$(2P). The pattern 
of differences in the 4th column agrees with expectations in 
the new scheme. All details concerning this calculation and 
results for some other quarkonia can be found elsewhere\cite{GlazekMlynik}. 
The oscillator frequency corresponding to Table \ref{Table} is 
$\omega$ = 182 MeV.
\begin{table}
\tbl{Qualitative illustration of bottomonium masses.}
{\begin{tabular}{@{}cccr@{}}\toprule
meson name      &  calculation (MeV)        & experiment (MeV)      & difference (MeV) \\
\colrule
$\Upsilon$10865 &             10725 &            10865 &-140 \\
$\Upsilon$10580 &             10464 &            10580 &-116 \\
$\Upsilon$3S    &             10382 &            10355&\hphantom{00}27 \\
$\chi_2$2P      &             10276 &            10269&\hphantom{000}7 \\
$\chi_1$2P      &             10256 &            10256&\hphantom{000}0 \\
$\chi_0$2P      &             10226 &            10232&\hphantom{00}-6 \\
$\Upsilon$2S    &             10012 &            10023&\hphantom{0}-11 \\
$\chi_2$1P&\hphantom{0}9912&\hphantom{0}9912&\hphantom{00}-1 \\
$\chi_1$1P&\hphantom{0}9893&\hphantom{0}9893&\hphantom{000}0 \\
$\chi_0$1P&\hphantom{0}9865&\hphantom{0}9859&\hphantom{000}5 \\
$\Upsilon$1S&\hphantom{0}9551&\hphantom{0}9460&\hphantom{00}91 \\
$\eta_b$1S&\hphantom{0}9510&\hphantom{0}9300&\hphantom{0}210 \\\botrule
\end{tabular}}
\label{Table}
\end{table}
The bottom line is that the realistic value of $\alpha$ is near 1/3 
and the new Hamiltonian approach to QCD can be further studied in 
a weak coupling expansion in the case of heavy quarkonia, including
many effects in the complex relativistic color dynamics of virtual 
quarks and gluons.

\end{document}